\documentclass[11pt]{article}

\usepackage{doublespace}
\usepackage{epsfig}
\usepackage{graphicx}
\usepackage{amssymb}
\usepackage{amsmath}

\begin{document}
\begin{spacing}{1.}
\title{The impressive complexity in the {\it Nautilus pompilius} shell}
\author{A. A. Castrej\'on Pita$^1$, J. R. Castrej\'on Pita$^1$,\\
A. Sarmiento Gal\'an$^2$, and R. Castrej\'on Garc{\'\i}a$^3$\\
\\
$1$ {\small Centro de Investigaci\'on en Energ{\'\i}a, UNAM, Ap. Postal 34,
62580 Temixco, Morelos, M\'exico.} \\
$2$ {\small Instituto de Matem\'aticas, UNAM, Av. Universidad s/n, 62200
Chamilpa, Morelos, M\'exico.} \\
$3$ {\small Instituto de Investigaciones El\'ectricas, Av. Reforma 113, 62490
Temixco, Morelos, M\'exico.}}
\maketitle
\end{spacing}

Running head: Impressive complexity in the {\it Nautilus pompilius} shell

\begin{spacing}{1.5}
\begin{abstract}
The complexity of the {\it Nautilus pompilius} shell is analyzed in terms of
its fractal dimension and its equiangular spiral form. Our findings assert 
that the shell is fractal from its birth and that its growth is dictated by a 
self-similar criterion (we obtain the fractal dimension of the shell as a 
function of time). The variables that have been used for the analysis show an 
exponential dependence on the number of chambers/age of the cephalopod, a 
property inherited from its form.

\end{abstract}
%\vfill\eject

\section{Introduction}

Fractal analysis is being applied with increasing frequency to living
organisms, trying to explain some of the complex forms found in nature. An
astonishing example reveals that {\it Ammonites} continuously increased their
complexity up to the point in which they became extinct \cite{Geology}. It is
our purpose to study in this paper the amazing complexity of a close relative
of the {\it Ammonites}, the {\it Nautilus pompilius.}

This pelagic species is a native of the western Indopacific ocean ($30^o$ N
lat. to $30^o$ S lat. and $90^o$ to $185^o$ W long. \cite{Saund87}), and
usually lives at a depth that varies from $50$ to $480$ meters (temperature
ranges from $24$ to $8$ Celsius degrees). The {\it Nautilus} reaches sexual
maturity at least $15$ years after hatching and then produces $10$-$15$ eggs
per year (it is not known if the female breeds more than once), and it may
live for $20$ years.

The shell is mother-of-pearl lined and pressure resistant (it implodes at
approximately $800 \ m$); its hardness has been the basis of various
ornamental handicrafts \cite{shells}. The vulnerability due to their slow
reproduction rate, and the fact that its exploitation has increased so much at
present, make it a possible new addition to the large list of endangered
species. But the most striking characteristic of this thin, two layered, and
spirally coiled shell is its internal subdivision in a series of successive
chambers (phragmocone), starting from the very moment of hatching when there
are already seven chambers present in the shell. As the cephalopod grows and
requires more space, it creates a new chamber by sealing the space behind it
with a calcareous septum and moves to live at the open, bigger end of the
shell. The rate at which a new chamber is created varies, at the beginning it
seems to take longer for the mollusc to seal the $8^{th}$ chamber but later
on, the process takes from $43$ to $77$ days per chamber \cite{Landman} and
lasts up to the completion of approximately $39$ sealed chambers \cite{Britannica} 
plus the open space where the mollusc lives \cite{Australia}; these changes in the
growth rate are easily understood in terms of the food availability and other
environmental variables. The sealing of the chambers however, is not complete,
there is a small duct in the center of each wall, called siphuncle, that
allows the living fossil to keep control of the pressure inside every previous
chamber and thus to regulate its buoyancy \cite{Guide}, \cite{Zoology}; the
heyday of the nautiluses is estimated to be around $500$ million years ago.

A transversal cut of the shell, Fig. 1, shows a perplexing spiral geometry, not
found in any other natural object; this is a black \& white image where the
borders have been prepared to facilitate the box-counting analysis. The
hemishell is $96.1 \times 106.2 \ mm$ and $32.2 \ mm$ wide; the number of
chambers is 30. Most amazing is the fact that its growth appears to be
self-similar, and thus for the shell to possess a fractal dimension. We now
proceed to confirm that this is indeed so.

\section{Method}

The digital image in Fig. 1 was obtained by placing half of the shell
directly on a scanner bed; the cutting was performed going through half of
the shell as accurately as possible. All measurements are performed on the
digital images, in pixel units, and the conversion factor is given by the
scanner resolution (72 pixels per inch). The borders of the edges in the
hemishell were previously tinted to gain contrast and improve definition,
and thus, making the contour threshold treatment unnecessary. It must be said
that we are assuming a perfect symmetry of the shell with respect to this cut.

As usually done when applying the box-counting method, it is necessary to
define a criterion for the size of the boxes, in particular for the maximum
possible size of the grid. This last value is easily determined: the use of
boxes bigger than the image size would produce constant values from there on
and thus a break down of the method. The box counting method is applied to the
original image and the fractal dimension of the whole shell is obtained via a
linear fit to the data \cite{Chaos},\cite{HaRFA}.

The previous selection however, means that the maximum possible size of the
grid will change when analyzing portions instead of the whole image; this is
precisely what happens if, in order to test the observed self-similarity, we
analyze the fractal dimension of smaller fragments of the image, that is, if
we check that its complex structure is the same regardless of the scale used
to measure it. To accomplish this test, we proceeded as follows; once the
box-counting method had been applied to the whole, bigger image, the last
chamber was digitally eliminated from the initial image and the method
reapplied to the new image after adjusting the maximum possible size to the
new, smaller image size, Fig. 2. This procedure was repeated up to the point
in which there were only the original seven chambers in the shell. The area of
the circumscribed rectangle was calculated for each step and the results are
shown in Fig. 3 in $mm^2$.

From the above procedure, we can also obtain the value of the different
intersections of the various straight lines with the vertical axis, {\it i.
e.}, the ordinates of each one of the lines obtained by the self-similar test.
Since these lines correspond to a different chamber number each, the result is
a function that can be used to predict the position and time of appearance of
the new chamber (and thus corroborate the average time mentioned earlier).

Now, according to the well known box-counting method, the fractal dimension
of an object, $D$, is defined as \cite{Chaos}:

\begin{equation}
D \ = \ \lim_{\varepsilon \rightarrow 0} \ \frac{ \ln N(\varepsilon)}{ \ln
1/\varepsilon},\label{dim}
\end{equation}
where $N(\varepsilon)$ is the number of boxes of a square grid of side-size
$\varepsilon $ required to cover the object in question. This definition comes
from the scaling law $N(\varepsilon) = C (1/\varepsilon)^D$, in which our
knowledge on integer-dimension objects is clearly expressed (one needs
$c/\varepsilon$ boxes of side-length $\varepsilon$ for each one of the $D$
dimensions of the object to be covered, where $c$ accounts for the `length'
in that dimension, and the $\varepsilon$-independent constant $C$ is merely
the product of the $c$'s). From this scaling law, a linear relation is
obtained:

\begin{equation}
\ln N(\varepsilon) = D \ln (1/\varepsilon) + \ln C,\label{linrel}
\end{equation}
and from it, definition (\ref{dim}) follows (the
$\varepsilon$-in\-de\-pen\-dent, constant term becomes negligible as
$\varepsilon \to 0$). We then see that the ordinate of the linear relation
(\ref{linrel}) can be interpreted as the natural logarithm of the number of
pixels of the original image (in practice, the grid can not be made smaller
than a pixel when $\varepsilon \to 0$).

To obtain adequate images for the use of the whole relation (\ref{linrel}) is
not an easy task, most of the known fractal objects do not have well defined
features like borders or surfaces \cite{Prasad}. In this case however, we have
all the images that were digitally generated to test the self-similarity of
the previously tinted shell (enabling us to avoid any contour threshold
analysis \cite{Prasad}), and thus all the data to build the graph in Fig. 4,
where we have plotted the image sizes (in $mm^2$) of the shell circumscribed by
a rectangle up to the next sealed chamber (Fig. 2) which correspond to the
ordinates in the linear relation (\ref{linrel}) (it is perhaps worth recalling
that the box-counting method only takes into account the pixels associated to
the contour of the shell when the size of the box is unity). We have also used
an average value for the time required for the construction of a new chamber in 
order to obtain the fractal dimension of the shell as a function of time, Fig. 5, 
this average value is $60 \pm 17$ days per chamber.

Having corroborated the fractal dimension of the shell at various scales and
the advantage of using the whole linear relation (\ref{linrel}), we can now
count the number of pixels (instead of boxes) inside each chamber; the results
are plotted in Fig. 6, where a linear relation is also clearly seen.

All these data enables us to predict the size of the new chambers and the time
of their appearance, that is, those chambers that would have formed in a
living specimen. The easiest way is to measure the volume of each chamber and
in order to achieve this, the hemishell was labelled on a Sartorius balance
($5 \ mg$ precision) and each chamber was filled up with water by depositing
one by one, $0.005 \ ml$ drops; the volume is then multiplied by two to
account for the other half of the shell. The internal surface of the shell
inhibits the formation of menisci and thus, the level of water inside each
shell is a flat surface that rises uniformly. The results are shown in Fig. 7.

As a final checking we fitted an equiangular spiral to the shell
\cite{Mathtabl}:
\begin{equation}
r \ = \ e^{\delta \theta },\label{spiral}
\end{equation}
where $(r,\theta)$ are the usual polar coordinates and $\delta$ is a parameter
that can be determined by the quotient of the shell distance from the center
in any direction and the same distance after a whole turn:
\begin{equation}
\ln \left( \frac{r_1}{r_2} \right) \ = \ 2 \pi \delta .\label{alpha}
\end{equation}
The value obtained from this quotient $\delta=(1+\sqrt{5})/2$, is the well known 
{\it golden ratio} and the resulting spiral is superimposed on the shell image 
and shown in Fig. 8; the spiral is represented by empty circles starting at the
closing wall of the eighth chamber.

\section{Results and Predictions}

The fractal (box-counting) dimension of the original {\it Nautilus} shell
shown in Fig. 1 is $1.635 \pm 0.006$; the average of the self-similar fractal
dimension of the shell (Fig. 5), obtained by the method exemplified in Fig. 2,
is $1.730 \pm 0.019$; this is an average over the life of the particular {\it
Nautilus} and clearly depends on the accuracy of the available data on the
shell growth. The lower value for fractal dimension of the original shell with
respect to the average, is due to the fact that the shell extension where the
mollusc lives is included in the original image (Fig. 1)

The other variables used in the preceding analysis show an exponential
dependence on the number of chambers or, equivalently, on the age of the
cephalopod:
\begin{equation}
y \ = \ \alpha_i \ e^{\beta_i x},\label{exprel}
\end{equation}
where $y$ is one of the properties described in Figs. 3, 4, 6, or 7,
$\alpha_i$ is the exponential of the ordinate in the linear relation shown in
figure $i$, $\beta_i$ is the slope in the corresponding relation (Fig. $i$),
and $x$ is the number of chambers in the image under analysis (or
equivalently, the age of the specimen). The relation obtained in Fig. 3 was
subsequently used as the criterion for the box-counting interval. The results
for the ordinates are: $\alpha_3 \ = \ 113.863 \ \pm  \ 0.039 \ mm^2, \ 
\alpha_4 \ = \ 46.016 \ \pm \ 0.039 \ mm^ 2, \ \alpha_6 \ = \ 4.852 \ \pm \ 
0.050 \ mm^2, \ $ and $ \ \alpha_7 \ = \ 0.030 \ \pm \ 0.072 \ ml$; for the 
slopes the values are: $ \ \beta_3 \ = \ 0.132 \ \pm \ 0.002, \ \beta_4 \ = 
\ 0.109 \ \pm \ 0.002, \ \beta_6 \ = \ 0.143 \ \pm \ 0.003, \ $ and $ \ 
\beta_7 \ = \ 0.192 \ \pm \ 0.003.\ $ The correlation factors for the linear 
relations in the above mentioned figures are: $ \ 0.996, \ 0.996, \ 0.992, \ $ 
and $ \ 0.997$ (in this case the linear fit was performed from the $11^{th}$ 
chamber onwards, see Fig. 7), respectively.

From our findings and within the shell growth accuracy, we can
also predict the appearance of a new chamber, the $31^{st}$ for
the analyzed specimen, with a volume of $ \ 11.347 \pm 0.010 \ ml,
\ $ or a transversal area of $ \ 415.71 \pm 0.005 \ mm^2 \ $ at $
\ 60 \ \pm \ 17 \ $ days after the last one had been completed.

\section{Conclusions}

In the previous analysis, we have shown that the shell of the {\it Nautilus
pompilius} that we have analyzed, possesses a frac\-tal dimension, that its
value is $1.635 \pm 0.006$ ($1.730 \pm 0.019$ on average), and that it does
not depend on the number of chambers (or, equivalently, the age) used to
calculate it. This establishes the self-similar structure of the shell at any
scale/time, and how its growth follows the same self-similar criterion. Hence,
we propose the measuring of the predicted appearance of a new chambers in
living {\it Nautilus}, even if only in laboratory specimens. Recalling that
the {\it Ammonites}, close relatives of the {\it Nautilus}, kept changing
their structure and disappeared from the face of the earth, one could also
conjecture that the preservation of the structure in the {\it Nautilus} has
meant an evolutive advantage for this species. This facet might help to the
conservation of this complex, living fossil.
\begin{spacing}{.5}
\footnote{This work has been partially supported by DGAPA-UNAM (IN101100),
CONACyT (32707-U), and UC-MEXUS.}
\end{spacing}

\end{spacing}

\vfill\eject

\begin{figure}[tbp]
\begin{center}
\epsfig{width=12.cm,file=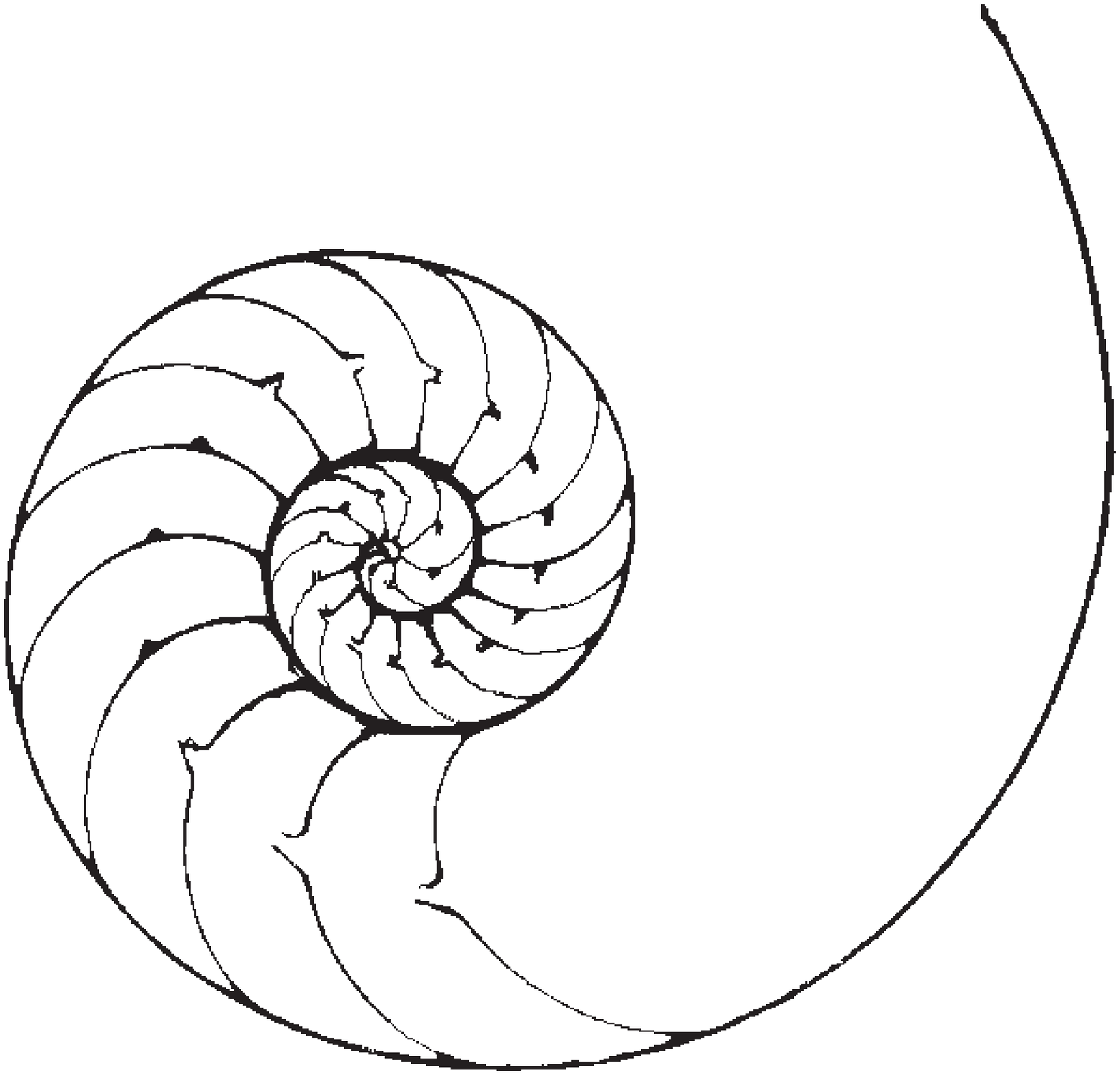}
\par
Fig. 1 Black and white image of a transversal cut of a {\it Nautilus
pompilius} shell (see text).
\end{center}
\end{figure}
\vfill\eject

\begin{figure}[tbp]
\begin{center}
\epsfig{width=14.cm,file=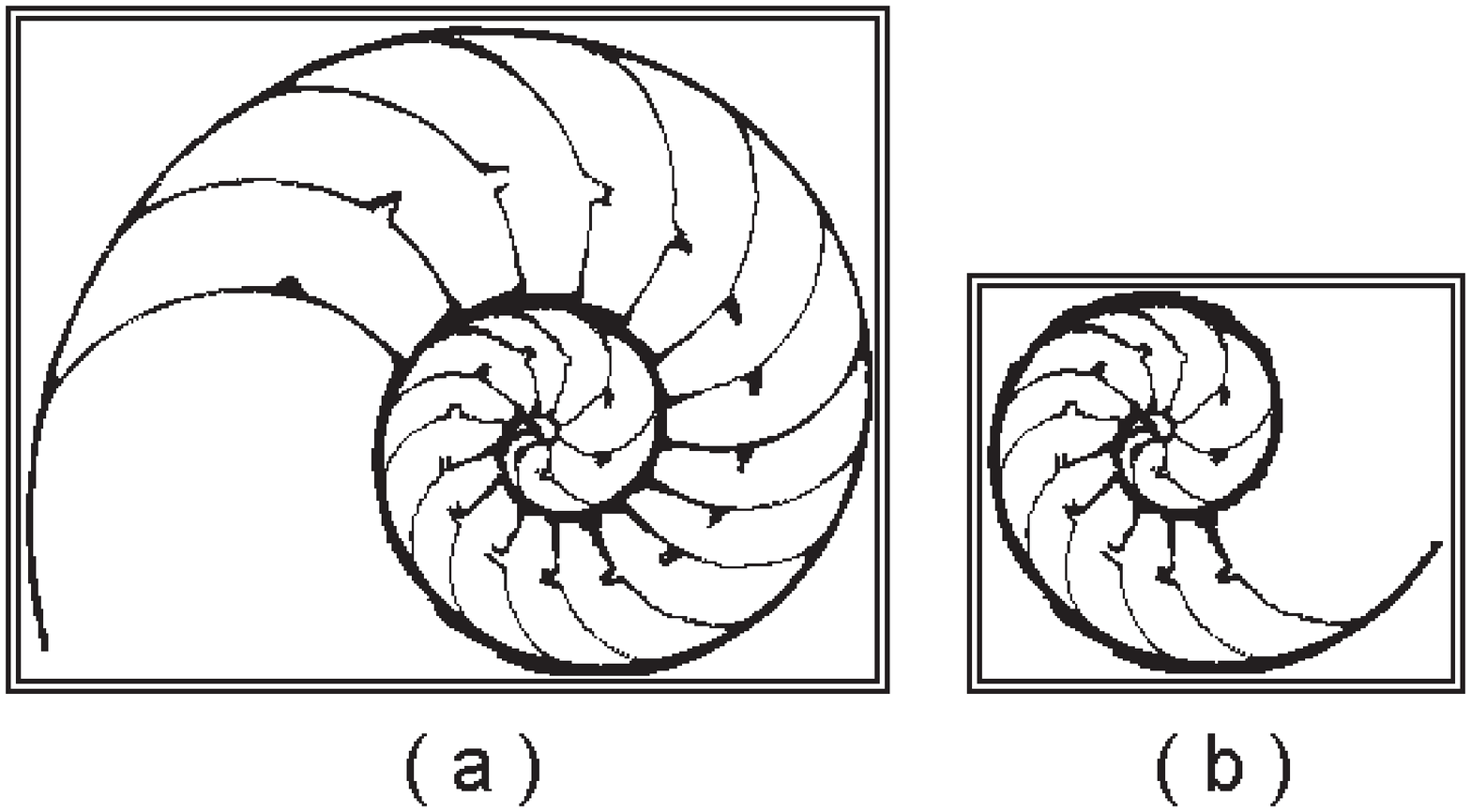}
\par
Fig. 2 Figure showing the images obtained by digitally altering the number of
chambers in the shell: (a) 25 chambers, and (b) 15 chambers.
\end{center}
\end{figure}
\vfill\eject

\begin{figure}[tbp]
\begin{center}
\epsfig{width=12.cm,file=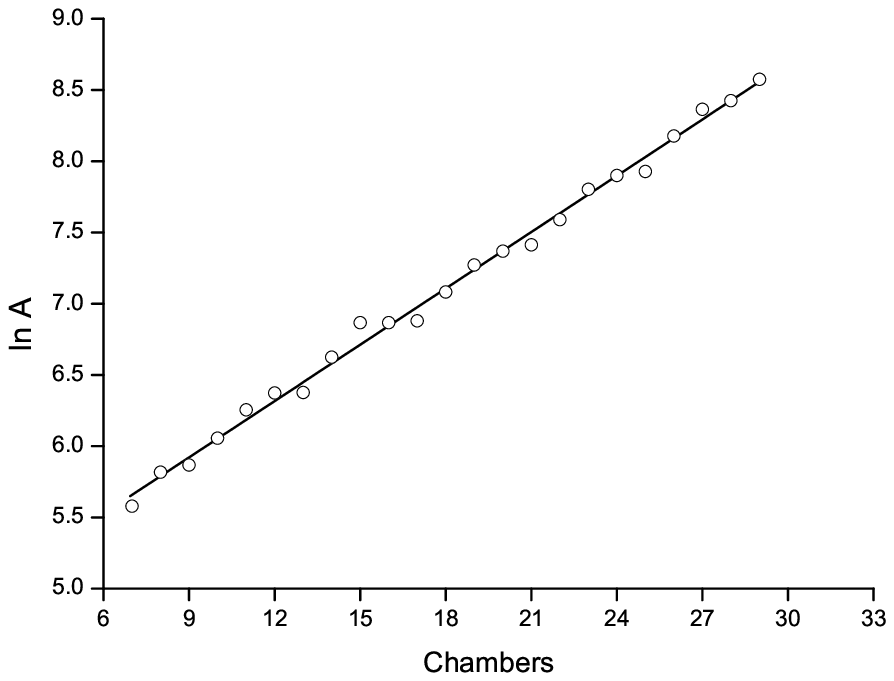}
\par
Fig. 3 Area ($A$) of the rectangle in which the {\it Nau\-ti\-lus} is
inscribed as a function of the number of chambers in the shell.
\end{center}
\end{figure}
\vfill\eject

\begin{figure}[tbp]
\begin{center}
\epsfig{width=12.cm,file=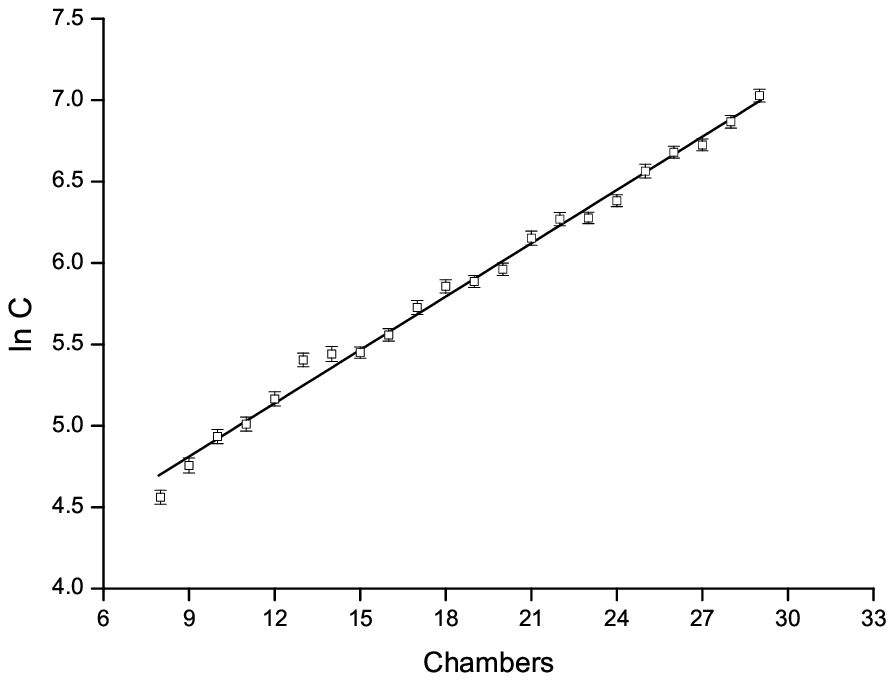}
\par
Fig. 4 Ordinates ($C$) of the lines obtained via the box-counting method as a
function of the number of chambers.
\end{center}
\end{figure}
\vfill\eject

\begin{figure}[tbp]
\begin{center}
\epsfig{width=12.cm,file=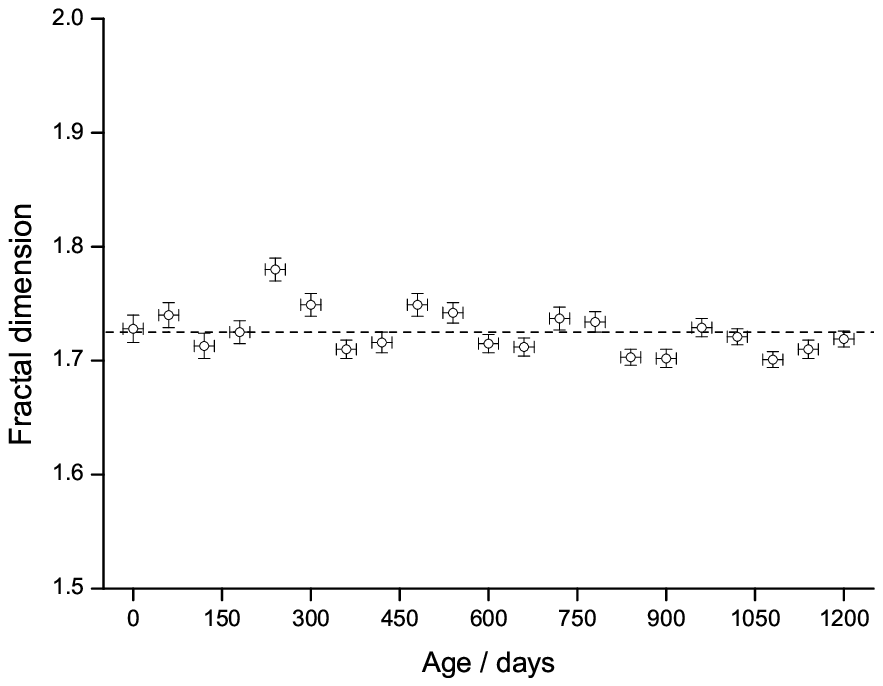}
\par
Fig. 5 Fractal dimension of the shell as a function of time, the age is 
measured in days after hatching, starting with the $8^{th}$ chamber. 
\end{center}
\end{figure}
\vfill\eject

\begin{figure}[tbp]
\begin{center}
\epsfig{width=12.cm,file=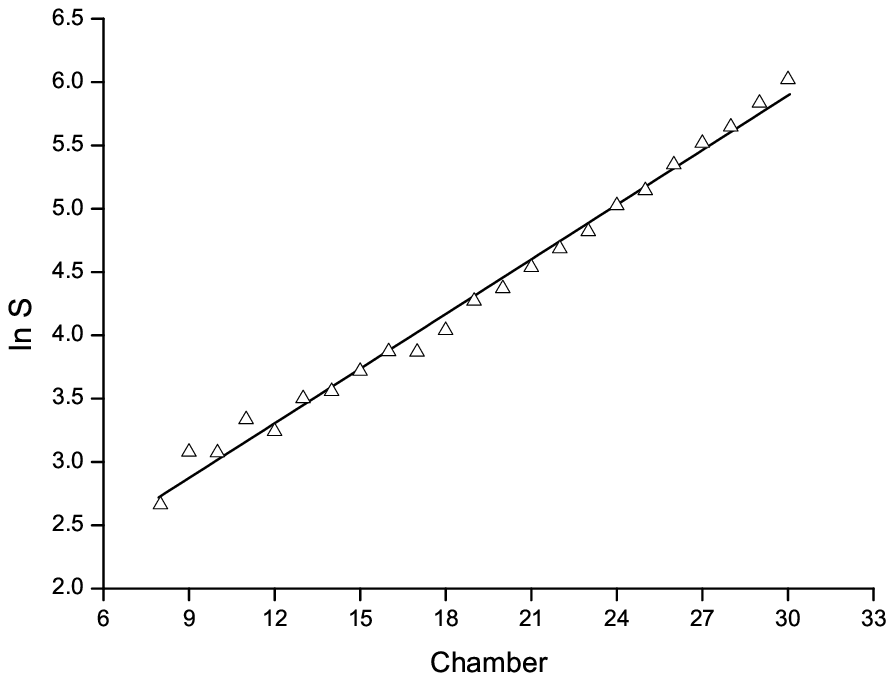}
\par
Fig. 6 Area of each chamber in pixels$^2$ ($S$) as a function of the number of
chambers.
\end{center}
\end{figure}
\vfill\eject

\begin{figure}[tbp]
\begin{center}
\epsfig{width=12.cm,file=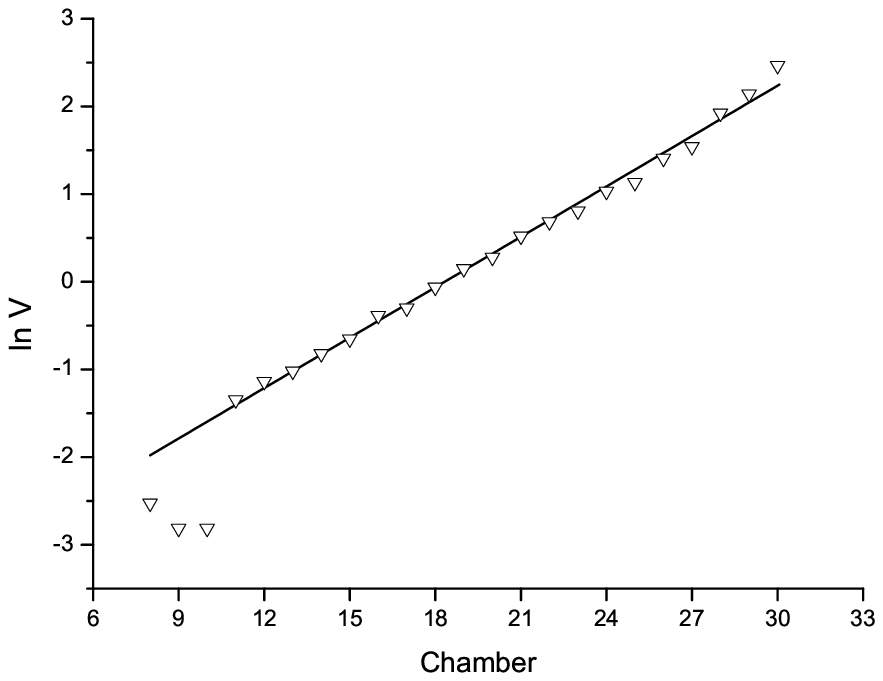}
\par
Fig. 7 Volume of each chamber in $ml$ ($V$) as a function of the number of
chambers.
\end{center}
\end{figure}
\vfill\eject

\begin{figure}[tbp]
\begin{center}
\epsfig{width=12.cm,file=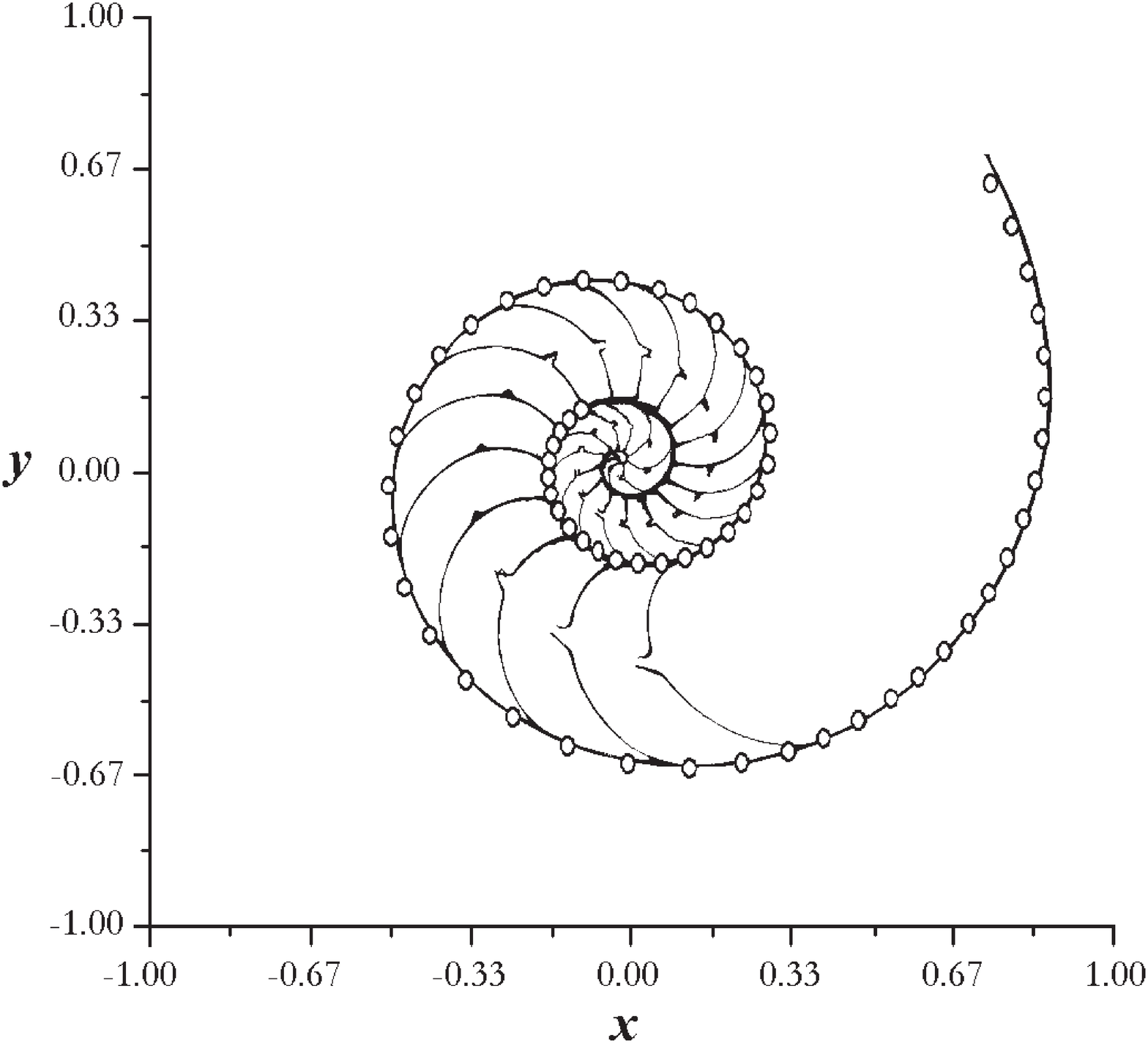}
\par
Fig. 8 Equiangular spiral (open circles) superimposed on the image of the {\it
Nautilus} shell.
\end{center}
\end{figure}

\end{document}